\documentstyle[graphics]{mn}

\newcommand{\etal}{{et al.}~}

\newcommand{\bc}{\begin{center}}
\newcommand{\be}{\begin{equation}}
\newcommand{\ee}{\end{equation}}
\newcommand{\ec}{\end{center}}

\newcommand{\hydra}{{\sc{HYDRA }}}

\newcommand{\cmbfast}{{\sc {CMBFAST }}}

\newcommand{\lya}{Ly$\alpha$~}

\newcommand{\ltsima}{\mbox{$\; \buildrel < \over \sim \;$}}
\def \simlt{\lower.5ex\hbox{\ltsima}}            
\def \gtsima{\mbox{$\; \buildrel > \over \sim \;$}}
\def \simgt{\lower.5ex\hbox{\gtsima}}            

\title[Wavelet analysis of QSO spectra] 
{A wavelet analysis of QSO spectra}
\author[Theuns \& Zaroubi]{Tom Theuns and Saleem Zaroubi\\
Max-Planck Institut f\"ur Astrophysik, Postfach 123, 85740 Garching, Germany}

\begin{document}
\maketitle

\begin{abstract}
The temperature of the intergalactic medium (IGM) is an important
factor in determining the line-widths of the absorption lines in the
\lya forest. We present a method to characterise the line-widths
distribution using a decomposition of a \lya spectrum in terms of
discrete wavelets. Such wavelets form an orthogonal basis so the
decomposition is unique. We demonstrate using hydrodynamic simulations
that the mean and dispersion of the wavelet amplitudes is strongly
correlated with both the temperature of the absorbing gas and its
dependence on the gas density. Since wavelets are also localised in
space, we are able to analyse the temperature distribution as a
function of position along the spectrum. We illustrate how this method
could be used to identify fluctuations in the IGM temperature which
might result from late reionization or local effects.
\end{abstract}


\section{Introduction}
Resonant absorption by neutral hydrogen in the intergalactic medium
along the line of sight to a distant quasar is responsible for the many
absorption lines seen in the \lya forest, blueward of the quasar's \lya
emission line (Bahcall \& Salpeter 1965, Gunn \& Peterson 1965; see
Rauch 1998 for a review).  The general properties of these \lya
absorption lines are remarkably well reproduced by hydrodynamic
simulations of cold dark matter (CDM) dominated cosmologies (Cen \etal
1994, Zhang, Anninos \& Norman 1995, Miralda-Escud\'e \etal 1996,
Hernquist \etal 1996, Wadsley \& Bond 1996, Zhang \etal 1997, Theuns
\etal 1998).

On large scales where pressure is unimportant, gas traces the dark
matter and the \lya spectrum can be used to infer the underlying
density perturbations in the dark matter (Croft \etal 1997, Nusser \&
Haehnelt 1999). On small scales however, pressure gradients oppose the
infall of gas into small potential wells (Jeans smoothing), leaving the
absorber more extended in space than the underlying dark matter. The
width of the absorption line is then determined by residual Hubble
expansion across the absorber (Hernquist \etal 1996), Jeans smoothing
and thermal broadening. Theuns, Schaye \& Haehnelt (2000) analysed
various line broadening mechanisms and demonstrated the importance of
the gas temperature in controlling the line-widths.

The strong dependence of the small-scale properties of the \lya forest
on the temperature of the gas allows one to reconstruct the thermal
evolution of the IGM. The gas temperature is set by the balance between
adiabatic cooling caused by expansion and photo-heating by the
UV-background. This introduces a tight relation between density and
temperature, $T=T_0(\rho/\langle\rho\rangle)^{\gamma-1}$ (Hui \& Gnedin
1997). The parameters $T_0$ and $\gamma$ of this \lq equation of
state\rq~ are very sensitive to the reionization {\em history} of the
IGM (Haehnelt \& Steinmetz 1998). This is because thermal time scales
are long in the low density IGM probed by the \lya forest, hence that
gas retains a memory about the past history of the ionising
background. Consequently, the \lya forest provides us with a fossil
record of the history of reionization, which can be explored by
unravelling its thermal history as deduced from the \lya forest.

Schaye \etal (1999, see also Ricotti, Gnedin \& Shull 2000) developed
and tested a method to infer $T_0$ and $\gamma$ based on the
line-widths of the absorption lines. Applying this method to high
resolution QSO spectra for a range of redshifts, they found (Schaye
\etal 2000) that the temperature $T_0$ decreases with decreasing
redshift as expected, however, there is a large increase in $T_0$ round
$z=3$, together with a decrease in the value of $\gamma$. They
attributed this change in the equation of state to late reionization of
helium II. They also noted that the temperature at higher redshifts is
still fairly high, which might be an indication that we are approaching
the epoch of hydrogen reionization.

The method of Schaye \etal to characterise line-widths is based on
Voigt profile fitting of absorption lines (Webb 1987, Carswell \etal
1987). The rationale behind fitting absorption lines with a Voigt
profile is partly historical, and stems from earlier theoretical models
in which the forest was produced by a set of \lya \lq clouds\rq~. The
line-width of these absorbers was assumed to be set by thermal and \lq
turbulent\rq~ broadening, which would produce a Voigt profile, and line
blending was responsible for the lines with large deviations from the
Voigt profile. In the new paradigm of the \lya forest absorption in the
general IGM is responsible for lines, and there is no a priori reason
to expect lines to have the Voigt shape.

In this paper we discuss a different method of characterising
line-widths, based on discrete wavelets (see e.g. Press et al. 1992 for
an introduction and further references). Wavelets provide an orthogonal
basis for a unique decomposition of a signal (the spectrum) in terms of
localised functions with a finite bandwidth. Thus they are a compromise
between characterising a signal in terms of its individual pixel values
and in terms of Fourier modes. In the first case, the characterisation
has no information on correlations between different pixels (no
frequency information) but perfect positional information. A Fourier
decomposition, on the other hand, has perfect frequency information but
no positional information. The analysis of a spectrum in terms of
wavelets has the advantage that one can study the clustering of lines
(\lq positional information\rq), as a function of their widths (\lq
frequency information\rq).

The usage of wavelets to analyse QSO spectra was pioneered by Pando \&
Fang (1996, 1998), who used a wavelet analysis of \lya absorption lines
to describe the clustering of those lines. The wavelet analysis
detected large scale structure in the \lya forest, which had proved
difficult using more traditional methods. In contrast to Pando \& Fang,
we will use wavelets to analyse the absorption spectrum directly,
thereby eliminating the somewhat subjective step of first decomposing
the continuous spectrum in absorption lines. The advantage of this new
method is that it allows us to objectively characterise the typical
width of absorption features as a function of position along the
spectrum\footnote{We will usually refer to absorption features as \lq
lines\rq, but this is just a convenient name for what the eye picks
out. The wavelet decomposition itself is unique and has no prejudice as
to what should be considered a line.}.

We will show using hydrodynamic simulations that the probability
distribution of wavelet amplitudes can be used to characterise the
equation of state of the absorbing medium, in terms of the temperature
at the mean density, $T_0$, and the slope, $\gamma$, of the
temperature-density relation. In addition we use the fact that wavelets
are localised in position along the spectrum, thereby allowing us to
detect spatial variations in $T_0$ and/or $\gamma$, which might be
present as a result of late helium II reionization or local effects.

This paper is organised as follows. In Section~\ref{sect:setup} we
first give a brief description of the generation of mock spectra from
our simulations and illustrate the decomposition of the spectra in
discrete wavelets. The statistics of the wavelet amplitudes for
different simulations is discussed in Section~\ref{sect:analysis} and
the results are summarised in Section~\ref{sect:conclusions}. Recently,
Meiksin (2000) discussed indepently the application of wavelets to QSO
spectra.

\section{Wavelet analysis of mock spectra}
\label{sect:setup}
\subsection{Mock spectra}
\label{sect:mock}
We use the L1 simulation described before in Theuns \etal
(2000). Briefly, this is a simulation of a flat, vacuum energy
dominated cold dark matter model with matter density $\Omega_m=0.3$,
baryon fraction $\Omega_b h^2=0.019$ and Hubble constant $H_0=65$ km
s$^{-1}$ Mpc$^{-1}$. Density fluctuations in this model are normalised
to the abundance of galaxy clusters (Eke \etal 1996) and we have used
\cmbfast (Seljak \& Zaldarriaga 1996) to compute the appropriate linear
transfer function. The IGM in this model is photo-ionised and
photo-heated by the UV-background from QSOs, as computed by Haardt \&
Madau (1996).

We simulated this cosmological model with a modified version of the
\hydra simulation code (Couchman \etal 1995), which combines
hierarchical P3M gravity (Couchman 1991) with smoothed particle
hydrodynamics (SPH, Lucy 1977, Gingold \& Monaghan 1977). We simulate a
periodic, cubic box of size 7.7 co-moving Mpc using 128$^3$ particles
of each species, which gives us sufficient resolution to compute
line-widths reliably (Theuns \etal 1998). To investigate other effects,
we also make use of simulations of a model with the same numerical
resolution, cosmology and thermal history, but with a smaller box size
(3.8 Mpc), and a set of simulations with a smaller normalisation
$\sigma_8=0.775$ and $\sigma_8=0.4$.

In the analysis stage, we impose a particular equation of state on the
gas at low overdensities ($\rho/\langle\rho\rangle < 20$) of the form
$T=T_0 (\rho/\langle\rho\rangle)^{\gamma-1}$, varying the values of
$T_0$ and $\gamma$. We then compute mock spectra that mimick the actual
observed HIRES spectrum of the $z_{\rm em}=3.0$ QSO 1107+485, discussed
by Rauch et al. (1997), using the following procedure. We divide the
observed spectrum in three redshifts bins, $z=2.5-2.625$,
$z=2.625-2.875$ and $z=2.875-3$ and scale the mean absorption of the
simulations at $z=2.5$, $z=2.75$ and $z=3$ to the corresponding
observed value. The simulated spectra are resampled to the observed
resolution, and convolved with a Gaussian to mimick instrumental
broadening. We have analysed the noise statistics of the QSO 1107
spectrum as a function of flux, and add noise with these properties to
the simulated spectra. By randomly combining individual sight lines
through the simulation volume, we generate a single long spectrum of
length 37 492 km s$^{-1}$.  Velocity $v$ is related to redshift $z$ via
$v\equiv c\left[log_e(1+z)-log_e(1+z_1)\right]$, where $c$ is the speed
of light, $z$ is redshift and $z_1$ is the redshift where \lya starts
to be confused with Ly$\beta$ for QSO 1107. In order to perform the
wavelet analysis, we resample the spectrum to $2^{15}$=32768 pixels,
equally spaced in velocity. In what follows, we will refer to a
simulation with a particular equation of state by giving $T_0/10^4$K
and $\gamma$, so the model $(1.5,5/3)$ has the imposed equation of
state $T=1.5\times 10^4\,(\rho/\langle\rho\rangle)^{2/3}$. We will
present results for four equations of state, using $T_0=1.5$ and
$2.2\times 10^4$K and $\gamma=1$ and 5/3.

\subsection{Wavelets}
\label{sect:wavelets}
\begin{figure*}
\setlength{\unitlength}{1cm} \centering
\begin{picture}(17,9)
\put(-1, -14.5){\includegraphics{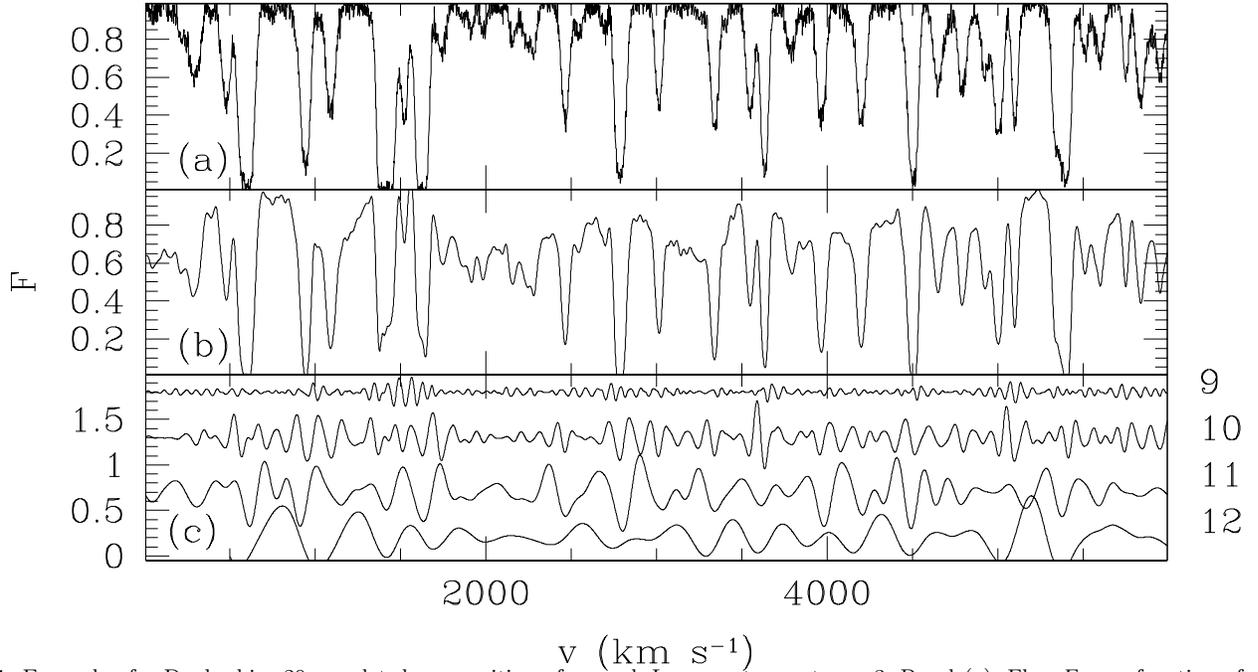}}
\end{picture}
\caption{Example of a Daubechies 20 wavelet decomposition of a mock
\lya spectrum at $z\sim 3$. Panel (a): Flux $F$ as a function of
velocity $v$ for a mock spectrum of QSO 1107. Panel (b): decomposition
of $F$ in terms of wavelets with resolutions $2^{i-15}\times V$ for
$i=9\cdots 12$ (from $18.3$ to 146.4 km s$^{-1}$) . Panel (c):
individual wavelets that make up the curve in (b), for $i=9$ (top
curve) to $i=12$ (bottom curve), off-set vertically for clarity. The
resolution corresponding to each wavelet is indicated on the right
axis. Most lines are detected in all shown wavelet resolutions, but
only narrow lines are strongly detected at the highest resolution
$i=9$.}
\label{fig:wltdecomp}
\end{figure*}

\begin{figure*}
\setlength{\unitlength}{1cm}
\centering
\begin{picture}(17,9)
\put(-1, -14.5){\includegraphics{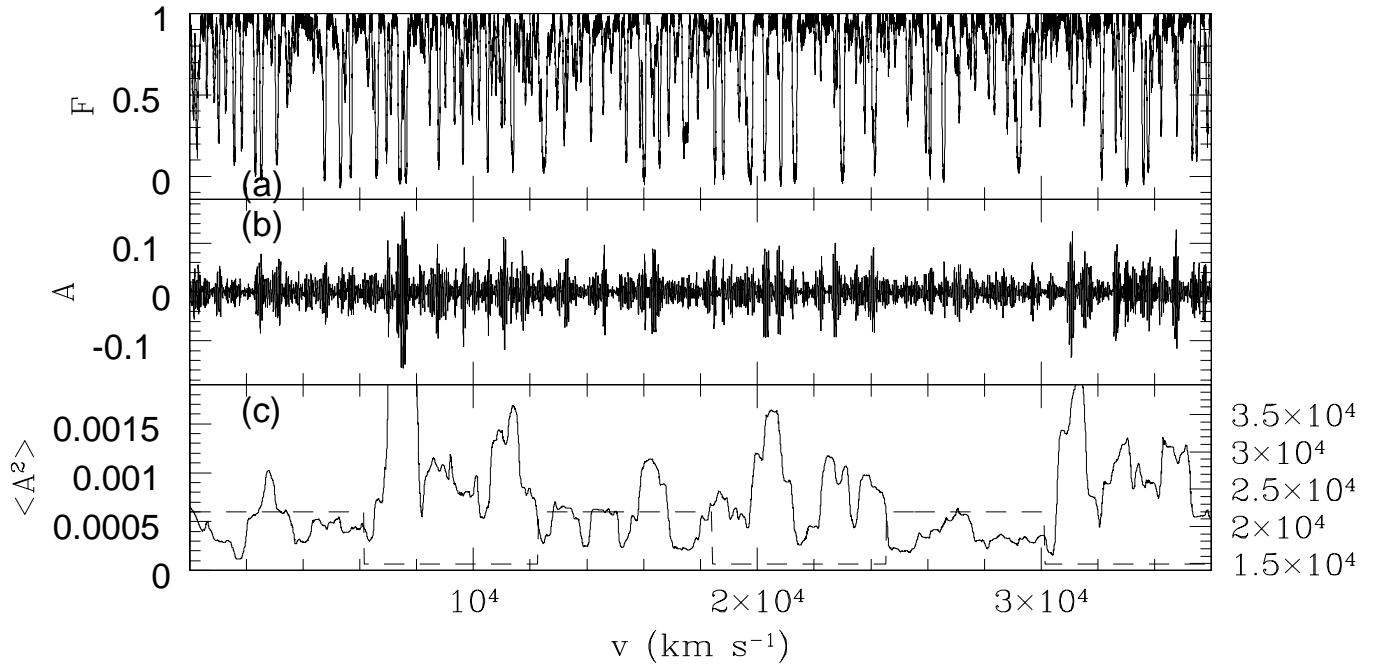}}
\end{picture}
\caption{Wavelet decomposition of a simulated spectrum at $z\sim 3$
(panel a) into the wavelet with resolution $i=9$ ($18.3$ km s$^{-1}$),
whose amplitude $A$ is shown in panel (b). The rms amplitude $\langle
A(9,1000)^2\rangle$, box-car smoothed over 1000 km s$^{-1}$ is shown in
panel (c) (full line). The simulated spectrum was made by combining
mock spectra from two models with different values of $T_0$ ($1.5\times
10^4$ and $2.2\times 10^4$K) but the same value of $\gamma=5/3$, in
stretches of length 6000 km s$^{-1}$. The temperature of this mixed
model is shown as the dashed line in panel (c) (right axis). There is a
strong correlation between the rms wavelet amplitude and the
temperature of the absorbing gas, with $\langle A^2\rangle$ on average
much larger for the cold parts of the spectrum, where $T_0=1.5\times
10^4$K, than in the hotter parts where $T_0=2.2\times 10^4$K.}
\label{fig:exmp}
\end{figure*}

The decomposition of a mock spectrum in terms of discrete wavelets is
unique, once a particular wavelet basis has been chosen. Here we will
use the Daubechies 20 wavelet (Daubechies 1988; see e.g. Press et
al. 1992 for a general discussion on wavelets, and an example of the
Daubechies 20 wavelet). Just as fast Fourier transforms, (discrete)
wavelets come in powers of two, but unlike Fourier modes, a given
wavelet has finite bandwidth and hence corresponds to a range of
frequencies. Nevertheless we will refer to a wavelet of a particular
\lq resolution\rq, for example quoting its full width at half
maximum. The simulated spectrum has a length of $V=$37492 km s$^{-1}$
and the wavelet resolutions correspond to $2^{i-15}\times V$. Here we
will use the exponent $i$ to refer to wavelets of a particular
resolution, e.g. $i=9$ corresponds to a wavelet of width 18.3 km
s$^{-1}$. Analysing a signal in terms of the amplitudes of wavelets
with different resolutions was pioneered in a different context by
Mallat (1989).

An example of a wavelet decomposition of a simulated spectrum is shown
in Figure~\ref{fig:wltdecomp}. Using wavelets with only four
resolutions ($i=9-12$) already gives a relatively good description of
the strong absorption features in the spectrum. Note how every line in
the top panel is \lq detected\rq~ on most resolution levels, indicating
that each individual absorption line is also made-up of a range of
frequencies.  This is of course because these lines are relatively well
approximated by Voigt profiles, which also have extended
bandwidth. However, some lines are only weakly detected in the $i=9$
narrow wavelet, while some of the narrower lines lead to large
amplitudes at this high resolution. It is this feature, namely that
some narrow lines are picked-up strongly by the narrow wavelets while
the broader lines are not, that allows us to characterise objectively
the typical line-widths of absorption lines.

For a smaller value $T_0$ of the IGM temperature, there will be a
larger fraction of narrow lines in the absorption spectrum. For a given
pixel at velocity $v$ in the spectrum, let
\begin{equation}
{\cal A}(v;i,W) \equiv \int_{v-W/2}^{v+W/2} A(v;i)^2
dv/W 
\end{equation}
denote the mean rms amplitude of the wavelet at resolution $i$, box-car
smoothed over a window of size $W$ (km s$^{-1}$). We will usually drop
the indices $i$ and $v$ in what follows, and assume $i=9$ unless stated
otherwise. For a spectrum with a larger fraction of narrow lines,
${\cal A}$ will be larger on average, hence we can in principle use the
statistics of ${\cal A}$ as a measure of $T_0$, once the relation
between them is calibrated with simulations.

In addition to this mean trend, ${\cal A}$ will fluctuate along the
spectrum, due to (random) fluctuations in the strengths of lines. Here
we give an example showing that averaging $A^2$ over a relatively short
stretch of spectrum is already enough to distinguish between models
with different $T_0$. This suggests it might be possible to detect {\em
fluctuations} in $T_0$ (and $\gamma$), which might be a relic of a
recent epoch of reionization or local effects. We will present a more
detailed analysis of how this can be done below and restrict ourselves
here to a typical example illustrated in Figure~\ref{fig:exmp}. To make
the shown spectrum, we have combined spectra of the $(1.5,5/3)$ model
on scales of 6000 km s$^{-1}$ with spectra of the 30 per cent hotter
model $(2.2,5/3)$, into one long spectrum of length $V$. (In what
follows, we will refer to this model as the mixed-temperature model.)
The rms amplitude ${\cal A}(v;9,1000)$ of the $i=9$ (18.3 km s$^{-1}$)
wavelet, smoothed on 1000 km s$^{-1}$, is sufficiently different
between these two equations of state that stretches of the colder model
can readily be distinguished from the hotter one as regions with larger
${\cal A}$.

In this example, both models have been scaled independently to have the
same mean optical depth, corresponding to the observed value for QSO
1107. In reality, regions of higher temperature would tend to have
smaller optical depth because of the $T^{-0.7}$ temperature dependence
of the recombination coefficient. This would tend to decrease the
amplitude of the wavelets in the hotter regions even more, making it
{\em easier} to distinguish between hot and cold regions.
 
\section{Wavelet statistics}
\label{sect:analysis}
\subsection{measuring the equation of state}
\begin{figure}
\setlength{\unitlength}{1cm} \centering
\begin{picture}(7,9)
\put(-2.4, -4){\includegraphics{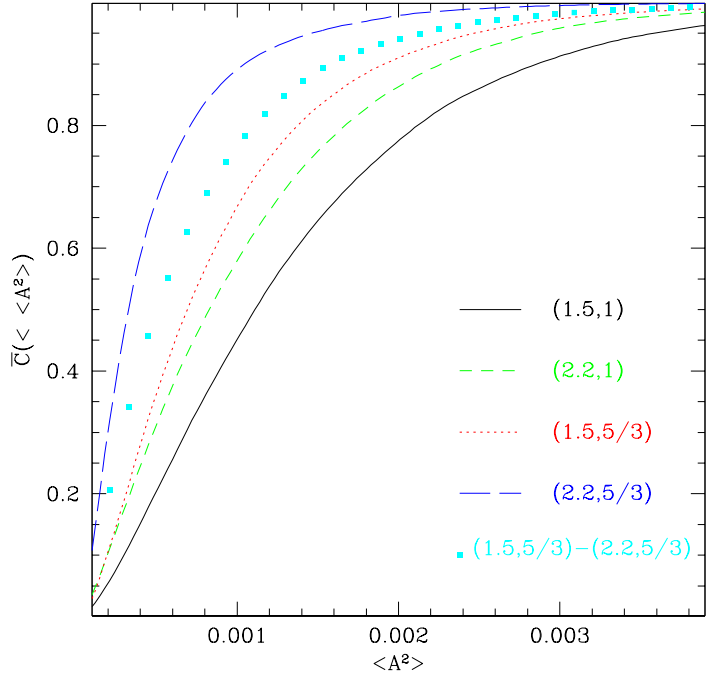}}
\end{picture}
\caption{Cumulative fraction $\bar C(< {\cal A})$ of pixels, where the
mean rms wavelet amplitude ${\cal A}\equiv \langle A(9,500)^2\rangle$
of the $i=9$ wavelet, box-car smoothed over a window of size $W=500$ km
s$^{-1}$, is less than some value, averaged over 100 spectra. The
different curves refer to different equations of state, as labelled in
the figure. Squares refer to the mixed-temperature model, obtained from
combining spectra of model $(1.5,5/3)$ with those of model $(2.2,5/3)$,
in stretches of length 6000 km s$^{-1}$. Models with smaller $T_0$ and
shallower equation of state have a larger fraction of pixels with large
values of ${\cal A}$. The mixed-temperature model differs from the
corresponding single temperature models.}
\label{fig:av500}
\end{figure}

\begin{figure}
\setlength{\unitlength}{1cm}
\centering
\begin{picture}(7,9)
\put(-2.5, -4){\includegraphics{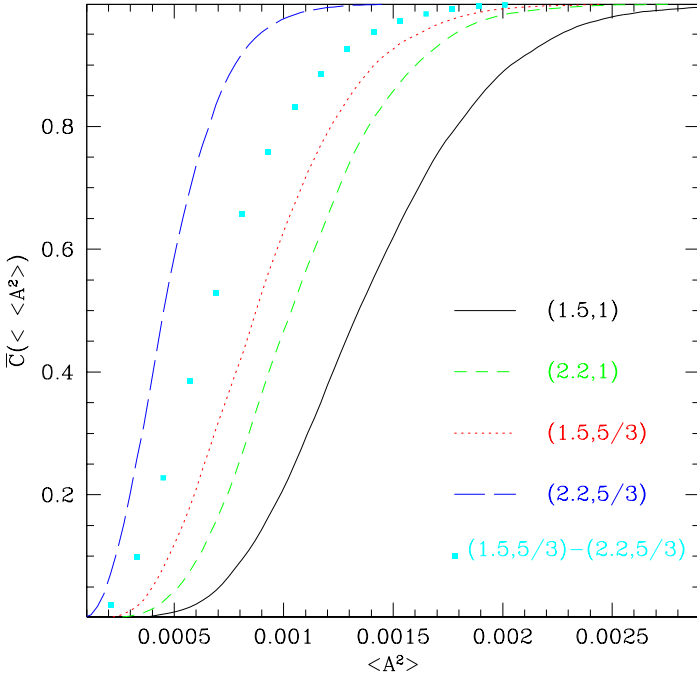}}
\end{picture}
\caption{Same as Figure~\ref{fig:av500}, but for a large smoothing
scale $W=2000$ km s$^{-1}$.}
\label{fig:av2000}
\end{figure}

\begin{figure}
\setlength{\unitlength}{1cm}
\centering
\begin{picture}(7,9)
\put(-2.5, -4){\includegraphics{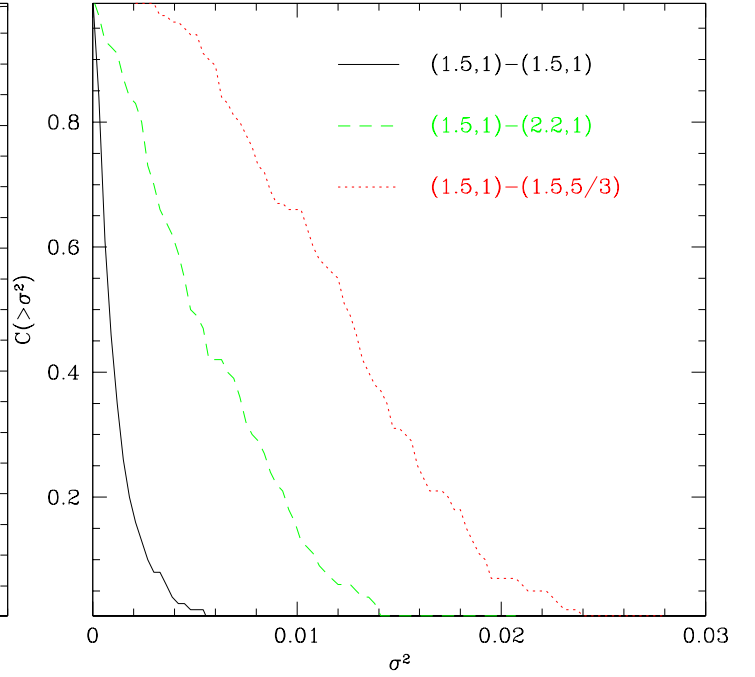}}
\end{picture}
\caption{Cumulative probability distribution of the dispersion
$\sigma_{ij}^2\equiv \langle (\bar C_i-C_j)^2\rangle$ for a smoothing
window $W=500$ km s$^{-1}$. The first model in the caption refers to
the model $i$ for which $\bar C_i$ is the mean (over 100 spectra)
cumulative distribution of ${\cal A}$. The second model in the caption
refers to model $j$ with cumulative distribution $C_j$ for a single
spectrum.  $\sigma_{ij}^2$ is a measure of the extent that model $j$
resembles model $i$. For $j=i$, it characterises the dispersion of the
cumulative distribution of model $i$ around its mean. The shown models
(2.2,1) and (1.5,5/3) can both be distinguished easily from model
(1.5,1) since $\sigma_{ij}^2$ tends to be $\ge 0.004$ for most
realisations of these models, whereas a realisation of model (1.5,1)
rarely deviates from its mean to such a large extent.}
\label{fig:disp500}
\end{figure}

\begin{figure}
\setlength{\unitlength}{1cm}
\centering
\begin{picture}(7,9)
\put(-2.5, -4){\includegraphics{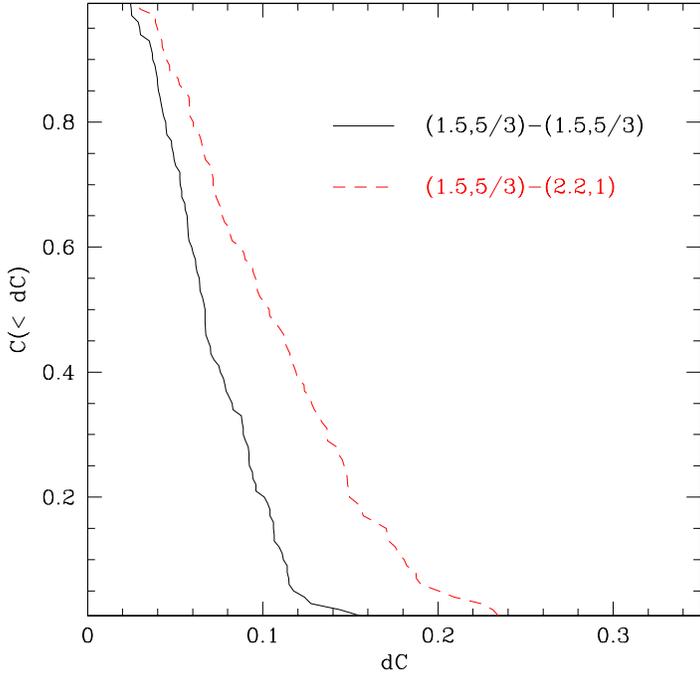}}
\end{picture}
\caption{Kolmogorov-Smirnov test whether a given realisation of model
$j$ (second model in the legend) is drawn from the distribution of
model $i$ (first model in the legend) for $i=(1.5,5/3)$ and $j=i$ (full
line) and $j=(2.2,1)$ (dashed line).}
\label{fig:kstest}
\end{figure}

\begin{figure}
\setlength{\unitlength}{1cm}
\centering
\begin{picture}(7,9)
\put(-2.5, -4){\includegraphics{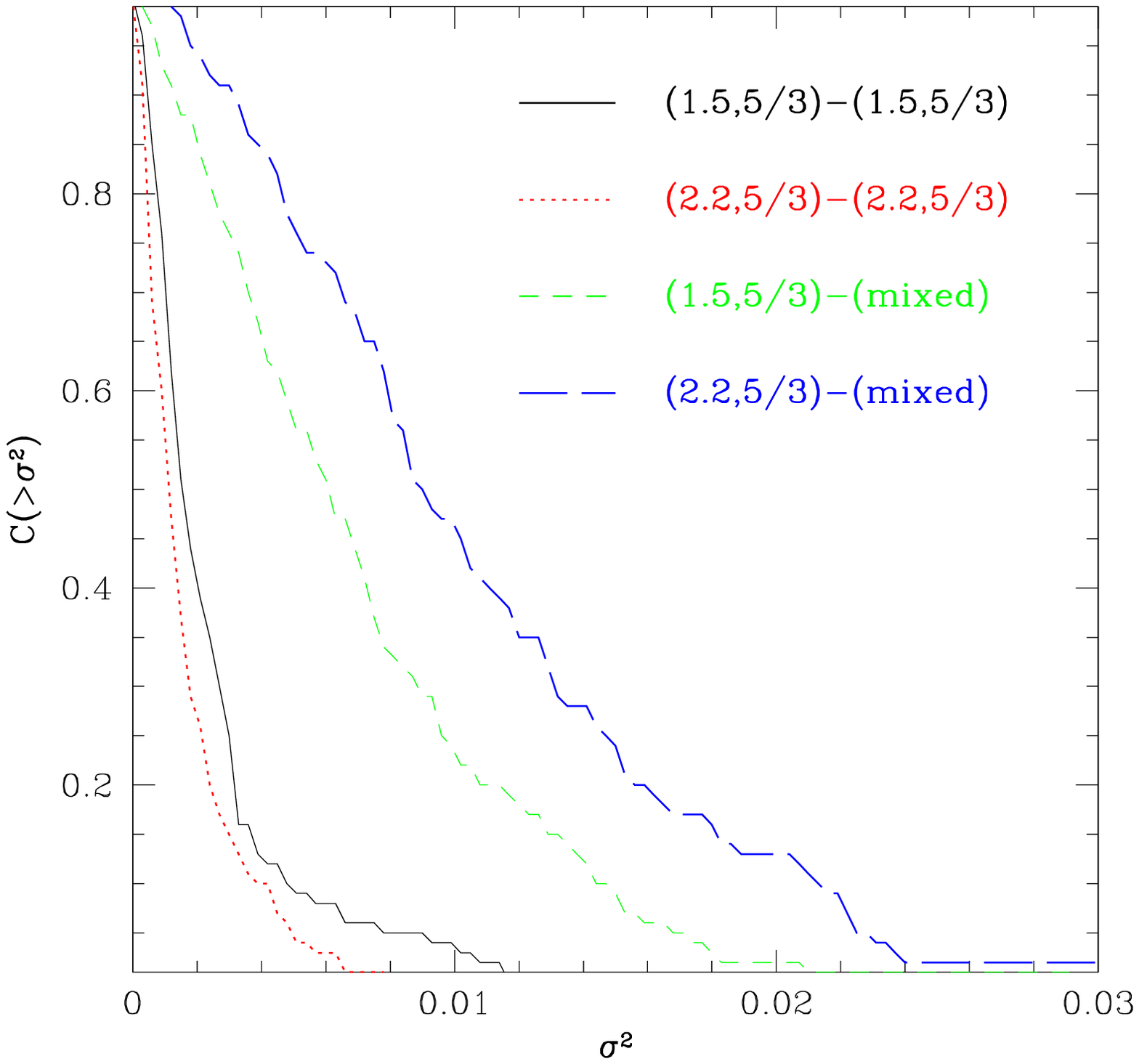}}
\end{picture}
\caption{Same as Figure~\ref{fig:disp500}, but for the
mixed-temperature model and with $W=2000$ km s$^{-1}$. The
mixed-temperature model resembles most strongly its cold constituent,
yet can be distinguished from it at the $>$ 80 per cent confidence
level.}
\label{fig:disp2000_mixed}
\end{figure}

In the previous section we showed that the rms amplitude of the $i=9$
narrow wavelet is strongly anti-correlated with the temperature of the
absorbing gas. Consequently we can characterise the temperature
distribution of the IGM over the spectrum using the corresponding
distribution of wavelet amplitudes. For each of 100 realisations of
models with a specified equation of state, we have computed the
cumulative distribution of ${\cal A}$,
\begin{equation}
C(< {\cal A}) = \int_0^{\cal A} P({\cal A}) d{\cal A}\,,
\end{equation}
where $P({\cal A})$ is the probability distribution of ${\cal A}$, and
we plot the mean over 100 realisations, $\bar C(< {\cal A})$, in
figures~\ref{fig:av500} and \ref{fig:av2000} for $W=500$ and 2000 km
s$^{-1}$, respectively.

As expected, the colder models are systematically shifted to larger
values of ${\cal A}$, since they contain a large number of narrow lines
and consequently have larger values of ${\cal A}$. Note, however, that
the dependence on the slope $\gamma$ is also quite strong, but this may
be partly a consequence of using the mean density as the pivot point
around which we change the slope. We have also superposed the
mixed-temperature model, which stays close to the hot component for
small values of ${\cal A}$ before veering away to the locus of the cold
component for large values of the amplitude.

Having shown that the mean cumulative distribution $\bar C({\cal A})$
depends on the equation of state, we now want to characterise how well
different models can be distinguished from each other, based on a {\em
single} spectrum. Hence, we want to characterise to what extent the
cumulative distribution $C_j({\cal A})$ for a single spectrum of model
$j$ differs from the mean, $\bar C_i$, for model $i$. To this end, we
compute the dispersion
\begin{equation}
\sigma_{ij}^2 \equiv \int_0^\infty (\bar C_i({\cal A})- C_j({\cal
A}))^2 d{\cal A}\,.
\end{equation}
For a single realisation of a spectrum of model $j$, $\sigma_{ij}^2$ is
just a number. In order to be able to distinguish between two models
$i$ and $j$ based on a single spectrum, it is necessary that the
dispersion $\sigma_{ii}^2$ be much smaller than the mean difference
$\sigma_{ij}^2$ between the models.

Figure~\ref{fig:disp500} shows the cumulative probability distribution
$C(>\sigma_{ij}^2)$ for $W=500$ for three different equations of
state. The confidence level at which a single spectrum of the model
with equation of state say (1.5,5/3) (model $j$) can be distinguished
from the model with equation of state (1.5,1) (model $i$) can be
directly read-off from this figure. For example, in $> 95$ per cent of
cases $\sigma_{ij}^2 > 0.004$ for $i\ne j$, whereas in only 2 per cent
of cases, a model which really has the equation of state (1.5,1) will
differ from the mean of this model to such a large extent.

A more usual statistic to judge whether a single realisation of a model
is drawn from a given probability distribution is the
Kolmogorov-Smirnov test, based on the maximum absolute difference
$dC={\rm max}|\bar C_i({\cal A}) - C_j({\cal A})|$ between two
cumulative distributions. The cumulative distribution of the
KS-statistic is shown in figure~{\ref{fig:kstest}}, where we compare it
for models (1.5,5/3) and (2.2,1), which resemble each other most in
figure~\ref{fig:disp500}. For 20 (5) per cent of realisations of model
(1.5,5/3), $dC>0.1$ ($dC>0.12$), and at this level of contamination, 60
(40) per cent of models (2.2,1) have $dC$ larger than that.

Finally, figure~\ref{fig:disp2000_mixed} illustrates how well the
mixed-temperature model can be distinguished from either the cold or
the hot model with $\gamma=5/3$. This model is most likely mistaken
with the colder single temperature counterpart. In 70 (25) per cent of
cases, the mixed model has $\sigma_{ij}^2 > 0.004$ ($\sigma_{ij}^2 >
0.01)$. This happens for the cold model in only 10 (5) per cent of
realisations.

\subsection{other effects}
\begin{figure}
\setlength{\unitlength}{1cm}
\centering
\begin{picture}(7,9)
\put(-2.5, -4){\includegraphics{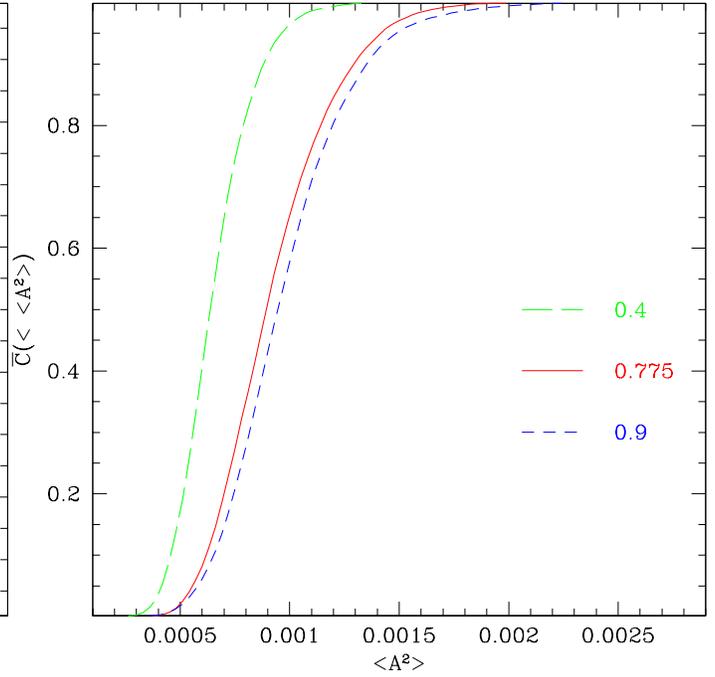}}
\end{picture}
\caption{Same as figure~\ref{fig:av2000} but for models with different
values of the normalisation $\sigma_8$, indicated in the legend. Models
with smaller $\sigma_8$ resemble more clustered but hotter models, but
the effect is relatively weak for realistic values of $\sigma_8$.}
\label{fig:sigma8}
\end{figure}

\begin{figure}
\setlength{\unitlength}{1cm}
\centering
\begin{picture}(7,9)
\put(-2.5, -4){\includegraphics{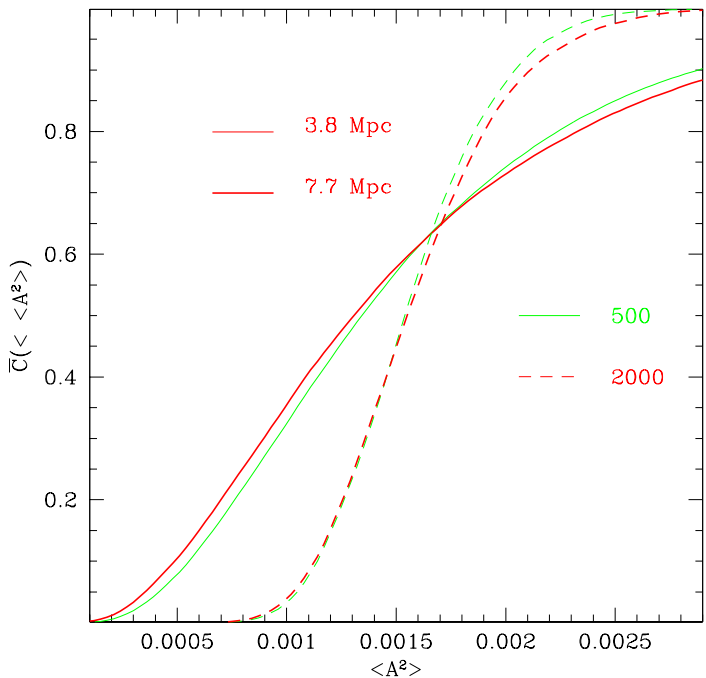}}
\end{picture}
\caption{Cumulative distribution $C(< {\cal A})$, for a smoothing scale
of 500 km s$^{-1}$ (full lines) and 2000 km s$^{-1}$ (dashed lines),
for a simulation box of 3.8 Mpc (thin lines) and 7.7Mpc (thick lines),
but the same numerical resolution. As before, we have plotted $C(<
{\cal A})$ averaged over 100 random realizations of the particular
model. The model shown is (1.5,5/3) and the spectra are scaled to a
mean effective optical depth of 0.26 at $z=3$. The influence of missing
long wavelength perturbations on the wavelet statistic is small.}
\label{fig:boxsize}
\end{figure}

Absorption features are broader in models with a smaller amplitude of
the dark matter fluctuations (Theuns et al. 2000), thereby resembling
more clustered but hotter models. This may lead to a degeneracy between
$T_0$ and $\sigma_8$ (Bryan \& Machacek 1999; note that Theuns, Schaye
\& Haehnelt (2000) showed that their Voigt profile analysis does not
suffer from such a degeneracy). For the statistic presented here, this
degeneracy is not very strong, as shown in figure~\ref{fig:sigma8}. The
model with $\sigma_8=0.775$ does not differ much from its more
clustered counterpart with $\sigma_8=0.9$. Only for very low levels of
clustering, $\sigma_8=0.4$, is the effect important. All models have
been scaled to a mean effective optical depth of 0.26 at a redshift
$z=3$.

Finally we have investigated the influence of the small box size in our
numerical simulations, and the result is shown in
figure~\ref{fig:boxsize}. Lack of long wavelength perturbations
decreases the observed range in ${\cal A}$, as expected, but the effect
of this purely numerical artifact is relatively weak.

\section{Conclusions}
\label{sect:conclusions}
Clues to the thermal history of the Universe are hidden in the small
scale structure of the \lya forest. There are two reasons for
this. Firstly, the widths of absorption lines are very sensitive to the
temperature of gas, and secondly, thermal time scales are long in the
low-density IGM that is responsible for the \lya forest. Since the
temperature of the photo-ionised IGM is determined by the evolution of
the ionising background, unravelling the thermal history will have the
added benefit of putting strong limits on the sources of UV light at
high redshifts.

We have presented a new way of analysing the small scale structure of
the \lya forest, based on the unique decomposition of a spectrum in
discrete wavelets. We have shown that the rms amplitude $\langle
A^2\rangle$ of narrow wavelets (18.3 km $^{-1}$) correlates strongly
with the temperature of the IGM, and also depends on the slope of the
equation of state. We have quantified to what extent different models
can be distinguished, using statistics of $\langle A^2\rangle$

Our mock spectra have been designed to mimick an observed spectrum of
QSO 1107+485 as much as possible. In particular, we have imposed on our
simulated spectra the same large scale optical depth fluctuations as
are observed in QSO 1107, making our mock spectra quite realistic. Even
so, we can still easily distinguish between models that differ in
temperature by less than 30 per cent. We have quantified the dependence
of these statistics on numerical artifacts (missing long wavelength
perturbations due to the smallness of our simulation box) and on the
amplitude of the dark matter fluctuations ($\sigma_8$).

Wavelets are also localised in space, making it possible to study $T_0$
and $\gamma$ as a function of position along the spectrum. We
characterised the extent to which we can distinguish models with a
single value of $T_0$ from a model with temperature fluctuations, as
might result from late reionization or local effects.

\section*{Acknowledgments}
We acknowledge simulating discussion with Martin Haehnelt, Michael
Rauch, Joop Schaye and Simon White, This work has been supported by the
\lq Formation and Evolution of Galaxies\rq~ network set up by the
European Commission under contract ERB FMRX-CT96086 of its TMR
programme.  This research was conducted in cooperation with Silicon
Graphics/Cray Research utilising the Origin 2000 super computer at
DAMTP, Cambridge.

{}
\end{document}